\documentclass[review]{elsarticle}
\usepackage{lineno,hyperref}
\usepackage{color}
\usepackage{xcolor}
\usepackage{booktabs}
\usepackage{threeparttable}
\usepackage{multirow}
\usepackage{float}
\usepackage{amsmath,amssymb,amsfonts}
\usepackage{array}
\usepackage{footmisc}
\usepackage{longtable}
\usepackage{booktabs}
\usepackage{threeparttablex}
\usepackage{marginnote}
\usepackage{appendix}

\newcommand{\question}[1]{{\textbf{#1}}}
\newcommand{\issue}[1]{{\textit{{#1}}}}

\newcommand{\finding}[1]{{\textbf{#1}}}
\newcommand{\DocumentationDebt}{{D}}
\newcommand{\CodeDebt}{{C}}
\newcommand{\DefectDebt}{{F}}
\newcommand{\TestDebt}{{T}}
\newcommand{\DesignDebt}{{S}}

\newcolumntype{L}[1]{>{\raggedright\arraybackslash}p{#1}}
\newcolumntype{C}[1]{>{\centering\arraybackslash}p{#1}}
\newcolumntype{R}[1]{>{\raggedleft\arraybackslash}p{#1}}

\journal{Information and Software Technology}

\begin{document}
\begin{frontmatter}

\title{Does it matter who pays back Technical Debt? An empirical study of self-fixed TD}

\author[firstaddress]{Jie Tan\corref{mycorrespondingauthor}}
\cortext[mycorrespondingauthor]{Corresponding author}
\ead{j.tan@rug.nl}

\author[firstaddress,secondaryaddress]{Daniel Feitosa}
\ead{d.feitosa@rug.nl}

\author[firstaddress]{Paris Avgeriou}
\ead{p.avgeriou@rug.nl}

\address[firstaddress]{Faculty of Science and Engineering, University of Groningen, Netherlands}
\address[secondaryaddress]{Campus Fryslân, University of Groningen, Netherlands}

\begin{abstract}

\textit{Context:} Technical Debt (TD) can be paid back either by those that incurred it or by others. We call the former self-fixed TD, and it can be particularly effective, as developers are experts in their own code and are well-suited to fix the corresponding TD issues. 

\textit{Objective:} The goal of our study is to investigate self-fixed technical debt, 
especially the extent in which TD is self-fixed, which types of TD are more likely to be self-fixed, whether the remediation time of self-fixed TD is shorter than non-self-fixed TD and how development behaviors are related to self-fixed TD. 

\textit{Method:} We report on an empirical study that analyzes the self-fixed issues of five types of TD (i.e., Code, Defect, Design, Documentation and Test), captured via static analysis, in more than 44,000 commits obtained from 20 Python and 16 Java projects of the Apache Software Foundation. 

\textit{Results:} The results show that about half of the fixed issues are self-fixed and that the likelihood of contained TD issues being self-fixed is negatively correlated with project size, the number of developers and total issues. Moreover, there is no significant difference of the survival time between self-fixed and non-self-fixed issues. Furthermore, developers are more keen to pay back their own TD when it is related to lower code level issues, e.g., Defect Debt and Code Debt. Finally, developers who are more dedicated to or knowledgeable about the project contribute to a higher chance of self-fixing TD.

\textit{Conclusions:} These results can benefit both researchers and practitioners by aiding the prioritization of TD remediation activities and refining strategies within development teams, and by informing the development of TD management tools.

\end{abstract}

\end{frontmatter}


\section{Introduction}
\label{sec:Introduction}

When Technical Debt (TD) results increasingly in extra effort during maintenance and evolution, it needs to be (at least partially) paid back, e.g., through refactoring. Previous studies have shown that a large percentage of source code Technical Debt (TD) is indeed paid back~\cite{DigkasSANER2018,RemovalICSME2017}. There is however, an elementary distinction between two cases of TD repayment, depending on whether the debt is paid back by the developer who introduced it or by others. We call the former \textit{self-fixed TD}, and it is significant, because developers are familiar with their own code, and may perform the repayment differently than others. On the contrary, we call the latter \textit{non-self-fixed TD}. Of course, remediation of TD incurred by others also takes place systematically. For example, developers that join an open source project are ideally pointed to easier tasks, which can involve update of documentation or simple code hacks~\cite{steinmacher19newcomers}. This helps the newcomers to build know-how and get used to the team's operations.

In spite of the intriguing nature of the mentioned distinction, the phenomenon of self-fixed technical debt has not been thoroughly studied so far. Investigating this can help identify good and bad practices in TD remediation, especially when comparing TD that is self-fixed and TD that is not. In particular, there are two aspects to study with regards to self-fixed TD: the phenomenon itself, and how human factors are related to it.

Studying the phenomenon itself may help to understand which types of TD are self-fixed and which ones are not. This can provide an insight on how developers prioritize the remediation of different types of TD, either their own or of others. Moreover, it may highlight cases where TD lives long in systems before it gets fixed and thus forces developers to pay high amounts of interest. Such findings can also indicate how developers generate value (measured as improved quality) by managing their own TD. For example, learning what types of TD developers are more keen to self-fix or to self-fix faster may hint towards practices to boost software value creation.

Studying human factors is also important as they can have a significant impact on software quality. For example, a strong feeling of ownership over a piece of code might increase the pride of accomplishment and contribute to delivering code with higher quality and fewer defects~\cite{Rahman_ICSE11}. Investigating such factors can further explain why developers self-fix TD differently, and why some projects are more likely to have TD that is not fixed. To the best of our knowledge, only a few studies have addressed the relationship between human factors and technical debt management~\cite{Amanatidis_XP2017,Alfayez_TechDebt18}; among them, there is no research that focuses on how human factors influence developers to self-fix TD. 

In this paper, we report on the extension of a previous study~\cite{tan2020empirical}, which is, to the best of our knowledge, the first study to exclusively focus on self-fixed technical debt. First, we investigate the extent of this phenomenon, as well as the types of TD that are more likely to be self-fixed and the survival time of self-fixed TD, especially compared to non-self-fixed TD. Second, we examine the relationship between human factors related to the effort of developers and development teams against the self-fixing rate, together with exploring the potential aspects that may influence developers in fixing their own TD. We emphasize that the extension comprises: a larger (approx. doubled) dataset including an additional programming language, namely Java; two new research questions regarding the examination of human-related factors; a more rigorous statistical analysis; and, finally, a revisit of all research questions to confirm previous findings.

To perform the study, we analyzed more than 17K commits from 20 Python projects and 27K commits from 16 Java projects of the Apache Software Foundation and investigated five types of TD, namely Code Debt, Defect Debt, Design Debt, Documentation Debt and Test Debt; these five types are defined by Alves et al.~\cite{ClassifyTD2014} and Li et al.~\cite{Li2015SMS}.
We used SonarQube 7.0 to detect and measure TD by identifying violations of a number of rules, which correspond to the different types of TD. 
The justification for the choice of projects, language and tool is elaborated in Sections \ref{subsec:Case Selection} and \ref{sec:variables}. 

The results indicate that the phenomenon of self-fixed TD is prevalent, as about half of the fixed issues are self-fixed. In addition, Defect Debt and Code Debt receive more attention from developers who introduced them, which may indicate that these developers are more keen to pay back their own TD when it is related to lower code level issues (e.g., compared to Design Debt and Documentation Debt). 
Moreover, there is no significant difference in the survival time between self-fixed and non-self-fixed issues.

Looking at project characteristics, the likelihood of TD issues to be self-fixed is negatively correlated with project size, the number of developers and total number of issues. Despite that, having developers who are more dedicated to or knowledgeable about the project (e.g., by contributing more commits) may positively impact the probability of the issues to be self-fixed. However, the longer the developers are involved in a project, the less often they self-fix TD.

Summarizing our findings, we have explored the phenomenon of self-fixed technical debt and how human factors are related to how developers self-fix technical debt. Broadly, the contribution encompasses the following items:
\begin{itemize}
    \item This is the first study to focus on how developers pay back their own technical debt in a comprehensive way, and looks at two of the most popular programming languages, i.e., Python and Java. 
    \item The study provides a comparison of the self-fixed and non-self-fixed TD items in terms of number, type of debt, and survival time.
    \item The study provides insight into the circumstances (e.g., project size and the number of developers) that would improve the likelihood of developers self-fixing technical debt, which can help predict project quality.
    \item The replication package of the study is publicly available, so other researchers can advance the work in the area of self-fixed technical debt.
\end{itemize}

The remainder of this paper is structured as follows.
Section~\ref{sec:Study_design} reports on the study objectives, the research questions, and provides details regarding the data collection and analysis. 
Section~\ref{sec:Results} illustrates the results of our study and Section~\ref{sec:Discussion} discusses the results in depth and their implications. 
Section~\ref{sec:Threats to Validity} reports on the threats to the validity of our study. 
Section~\ref{sec:Associated Dataset and Replication} provides information about our dataset and replication package. 
After the discussion of related work in Section~\ref{sec:Related_work}, Section~\ref{sec:Conclusion and Future Work} concludes the paper and outlines the directions of future work.

\section{Related work}
\label{sec:Related_work}

The body of knowledge in TD management has grown substantially over the years. We are particularly interested in work pertaining to TD remediation and human factors related to it. 

Significant research on TD remediation has focused on fixing code smells, which is a very popular sub-type of design debt~\cite{Li2015SMS}. In this context, Chatzigeorgiou et al.~\cite{Chatzigeorgiou_Evolution_Bed_Smells} found that the amount of code smells increases over time and Tufano et al.~\cite{TSE2017} that some smells could survive for a long time. 
Focusing on maintenance activities, Palomba et al.~\cite{Impact_ESE2018} found that 
the removal of code smells is beneficial for code change-proneness most of the times. Sj\o{}berg et al.~\cite{maintenance_TSE} identified that the 12 analyzed code smells have no significant effects on maintenance effort. Finally, Digkas et al.~\cite{DigkasSANER2018} observed that a large percentage of TD, including code smells, is paid back during software evolution. 

Another aspect of TD remediation regards the concept of self-admitted technical debt (SATD), i.e.~developers admitting incurring TD (e.g. in source code comments). 
In this context, Potdar and Shihab~\cite{SATDICSME2014} extracted SATD based on source code comments and found that developers introduce and remove SATD throughout all development activities, i.e., they do not only introduce or remove SATD during the beginning or end of their development activities. 
Other studies found that the majority of SATD is removed and often by the same developer who introduced it~\cite{SATDMSR2016,RemovalICSME2017}.
However, Potdar and Shihab~\cite{SATDICSME2014} also noticed that developers with more experience tend to introduce more SATD. In Section~\ref{sec:DiscussionHF}, after presenting our results, we discuss how these observations compare to our findings. Furthermore, Wen et al.~\cite{wen_2020ICPC} manually analyzed 500 quick remedy commits, i.e., the commits that fix issues introduced by the same developers as the result of omitted code changes or errors in the previous commit. Such commits were identified by looking at their objective. During the analysis, Wen et al. found 49 remedy commits related to software documentation, among which 16 are related to code comment (including missed removal of SATD instances).

Unlike the studies mentioned in the previous paragraphs, which considered one or two types of source code technical debt, or SATD, we analyzed five common types: Code, Defect, Design, Documentation and Test Debt in source code. Among these five types of TD, one previous study has shown that developers mostly considered Design Debt, Test Debt and Defect Debt~\cite{codabux_2013MTD}. Regarding TD remediation, we focused on who fixed TD, i.e., whether it was fixed by the developer who introduced it or not. In particular, we investigated the types and survival time of TD in a comparison between self-fixed and non-self-fixed TD.

Since software development is the result of teamwork, factors related to the development team can have a significant impact on software quality. For example, casual developers are likely to introduce quality-related problems of greater severity and in greater quantity~\cite{Lu_APSEC2016}. 
Thus, investigating the factors related to developers and development team may further explain why the self-fixing rates vary among different projects. In our study, such factors include the number of developers and the frequency of their modifications.

Regarding the relationship between TD and human factors, Amanatidis et al.~\cite{Amanatidis_XP2017} investigated to what extent seniority relates to a tendency to accumulate more TD. Although they found a negative association, the evidence is still inconclusive.
In addition, they discovered that 80\% of the most TD-incurring developers have low project-related experience. 
In a separate study, Alfayez et al.~\cite{Alfayez_TechDebt18} investigated how different developers and developer characteristics, such as developer's seniority, are related to the introduction of TD in 38 Apache Java systems.
They found that developer's seniority are negatively correlated with the amount of introduced TD. 
In contrast to these studies, we focused on remediation of TD rather than its introduction, and in the context of self-fixed TD.

There are also external factors that can affect how developers remediate technical debt, such as using tools. For example, Gilson et al.~\cite{Gilson_TechDebt2020} found that junior developers tend to use static analysis tools to value composite quality indicators (e.g., maintainability, reliability in SonarQube), even if they do not fully understand their meaning. However, Ernst et al.~\cite{Ernst_FSE2014} conducted a survey with 1,831 participants, primarily software engineers and architects, and found that the lack of tool support for accurately managing and tracking architectural sources of debt is a key issue and remains a gap in practice. Thus, the findings of how developers remediate their own TD can be used to improve the current tools.

\section{Study design}
\label{sec:Study_design}

This empirical study was designed
based on the guidelines of Runeson et al.~\cite{Runeson:12} 
and is reported according to 
the Linear Analytic Structure~\cite{Runeson:12}.

\subsection{Objective and Research Questions}
\label{subsec:Research Questions}

The goal of our study is to investigate self-fixed technical debt with respect to the extent of this phenomenon, its different types and survival time, and its association with human factors by analyzing Python and Java projects.

Based on this goal, this study will examine the following five research questions:

\begin{description}
\item [\textbf{RQ1:}] \question{How much technical debt is self-fixed in open source projects?}

Ideally, a developer who introduces technical debt to source code is also the best candidate to pay it back in the future. Since developers have the best understanding of their own code, they are more likely to know the causes of technical debt they incurred and have potential remediation strategies. This research question aims to investigate to what extent developers indeed take care of their own technical debt and also investigate how project characteristics, i.e., SLOC, the number of commits and developers, and the total number of issues, relate to the likelihood that the contained issues to be self-fixed. Such findings can be used to further explain how developers pay back TD in projects with different characteristics, e.g., learning whether they are more likely to fix a TD item by themselves or wait for other developers to address it.

\item [\textbf{RQ2:}]\question{Which type of technical debt is more likely to be self-fixed?}

Different issues belong to different types of technical debt and require different amounts of effort to be resolved. Therefore, this research question examines which issues and their corresponding TD types are more likely to be fixed by the developers who introduced them.
The answer to this research question will shed  further light on how developers consider their own technical debt by allowing to look at how the type of TD plays a role on what extent TD is fixed.
For example, some issues may be associated with higher risks and thus be regarded as important to fix by the developers that introduced them. Or the opposite, some issues may be regarded as `low hanging fruits', i.e. easy to self-fix. In addition, software components containing issues that are mostly fixed by others could be inspected to ensure proper documentation to support future maintenance activities. 

\item [\textbf{RQ3:}]\question{How long does self-fixed technical debt survive during the evolution of a system?}

This research question focuses on investigating if self-fixed issues are addressed faster than the issues that are introduced by other developers, and comparing which types of TD issues are likely to be self-fixed in a shorter time.
Analyzing the survival time of self-fixed technical debt during the evolution of each system helps us better understand how developers prioritize technical debt repayment, and how much interest is paid accordingly.
Since developers tend to be more familiar with their own code, one would expect that developers can address their own TD issues in a shorter time. 

\item [\textbf{RQ4:}]\question{How does the number of developers and the frequency of their modifications relate with the likelihood of TD issues to be self-fixed?}
    
Previous research has shown that not only human aspects related to individual developers, but also the collaboration of team members can have a significant impact on software quality~\cite{Lu_APSEC2016}. Similarly, we explore not only the impact of individual developer behavior on the likelihood that the issues to be self-fixed, but also team behavior. In particular, we look into how many developers maintain a given file and how many times a file has been modified by all the participating developers.
In the former case, a higher number of developers maintaining a file/module at the same time (assuming that other factors are constant) may be associated with a higher chance of issues being fixed by developers that did not introduce them in the first place. In the latter case, the more a file changes, the more likely an issue is to be self-fixed.

\item [\textbf{RQ5:}]\question{Who (self) fixes technical debt?}

Developers have different ranges of involvement and contribution within a system. In general, developers with a high involvement are more familiar and experienced with their projects~\cite{Rahman_ICSE11}. This research question aims at investigating the possible factors that drive some developers to have a higher chance to self-fix TD than others, by looking at \textit{seniority} (i.e., how long they have been working on the project), \textit{file count} (i.e., how many files they have involved in), \textit{commit count} (i.e., how many commits they have authored) and \textit{commit size} (i.e., how many lines of code (LOC) they have contributed on average per commit).

\end{description}

\subsection{Case Selection}
\label{subsec:Case Selection}

For the purpose of this study, we selected Python and Java projects from the Apache Software Foundation (hereafter referred to as the Apache ecosystem) as subject systems. 
Python and Java are currently ranked as two of the top three most popular programming languages\footnote{According to the Tiobe Index, one of the best known indices of programming languages popularity, \url{https://www.tiobe.com/tiobe-index/}}. While our earlier study~\cite{tan2020empirical} solely focused on projects written in Python, we expanded this study to include Java projects, improving the external validity of our findings.

The Apache ecosystem is one of the largest open source foundation, with more than 8,000 Apache code committers\footnote{https://www.apache.org/, visited in November 2020}. 
In addition, the ecosystem has 55 Python and 1,000 Java projects on GitHub, which contain different domains, sizes, activity and number of files. 
Finally, Apache projects have long-term stability\footnote{http://www.apache.org/foundation/how-it-works.html, visited in November 2020}: 
a team of self-selected technical experts manages each project by following Apache-wide meritocratic rules, and an incubator filters projects based on their likelihood of becoming a successful community.

To select systems among the Apache projects on GitHub, we used four inclusion criteria:
\begin{enumerate}
    \item The project must show up on the Apache Projects List\footnote{\url{https://www.apache.org/index.html\#projects-list}, visited in September 2020}, which excludes Apache Incubator projects. 
    Incubated projects are on a transition period to conform to Apache standards and are, therefore, non-representative.
    \item  The project's main programming language must be Python or Java, i.e., the largest number of files and source lines of code (SLOC) are written in Python or Java.
    \item The project must involve at least two developers.
    \item The project must have at least 10\% of its issues fixed. A fixing rate lower than 10\% may indicate poor quality and decreased attention on maintenance.
\end{enumerate}

Analyzing the entire history of commits for some projects is prohibitive in terms of the required computational power and time.
Thus, to still guarantee the inclusion of sufficient revisions for long-lived software systems, we decided to use at least the first 2,000 commits for each project. We particularly focus on the first 2,000 (rather than for example the last 2,000) since it facilitates determining the authors of commits that introduce TD. 
For projects that have less than 2,000 commits, we considered their entire history.

Based on the criteria, we selected all Python projects that fit these criteria at the time of data collection (i.e., 20 projects). Compared to Python the number of potential Java projects is vast; thus, we randomly selected 16 projects. 
The selected projects partially represent the diversity of the Apache ecosystem in terms of SLOC, number of commits and developers.
We note that although one Python project has two developers, 17 projects have at least five developers (median: 9; average: 18; max 85). Moreover, among the 16 Java projects 13 have at least 20 developers (min: 15; median: 55; average: 89 max: 319). 
Also, the majority of the projects have a long history of commits (at least 3 years), which allows to investigate the survival time of self-fixed technical debt over an extensive period~\cite{Marinescu2012}.

\subsection{Variables and Data Collection}
\label{sec:variables}

In this section, we describe the sets of variables necessary to answer the research questions, as well as the necessary tooling and the major steps of the data collection process. 
In particular, each unit of analysis comprises the tuple:

\vspace{0.2cm}
\noindent{$<$\textit{snapshot identification}; \textit{snapshot information}; \textit{TD issues}$>$}
\vspace{0.2cm}

The \textit{snapshot identification} comprises the \textit{project name} and \textit{commit hashcode}.
The \textit{snapshot information} encompasses the \textit{developer's name},\textit{developer's email}\footnote{For privacy reasons, we do not disclose the developers' name and email in our replication package},  \textit{time-stamp} (i.e., date of commit), and list of files (added, removed and modified).
\textit{TD issues} concern the amount of self-fixed and non-self-fixed issues in a particular snapshot. Data collection is comprised of three main steps.

\vspace{0.2cm}
\noindent
\textbf{Step 1: Technical Debt Detection}

To perform our study, we use the tool SonarQube~\cite{SonarQube} to detect technical debt. 
There are two main reasons for choosing SonarQube: 
(1) it can track the evolution of technical debt by analyzing multiple versions of projects;
(2) it is being widely used in both industry~\footnote{\url{https://www.sonarsource.com/customers/}, visited in September 2020} and for research purposes~\cite{DigkasSANER2018,SonarICPC,SonarTechDebt,SQALE4,SQALE3}.
SonarQube defines a set of rules to detect various types of technical debt and classifies them into four severity levels: \textit{blocker}, \textit{critical}, \textit{major} and \textit{minor}.  
We do not consider \textit{minor} issues, since many of them, e.g., \issue{Lines should not end with trailing whitespaces}, are trivial and have low impact and likelihood\footnote{\url{https://docs.sonarqube.org/latest/user-guide/rules/}, visited in September 2020}. 
Moreover, some developers might not treat issues of \textit{minor} severity as technical debt~\cite{DigkasSANER2018}.
During the analysis, SonarQube creates a new issue when a piece of code breaks one of the predefined rules and also assigns a time estimate of how long it would take to resolve it.

To simplify the organization of the technical debt issues, we grouped the rules into five categories, i.e., Code Debt, Defect Debt, Design Debt, Documentation Debt and Test Debt, according to the classification given by Alves et al.~\cite{ClassifyTD2014} and Li et al.~\cite{Li2015SMS}. In addition, we also filtered out rules that could not be meaningfully or directly mapped into one of the studied TD types. For example, we did not consider the FIXME rule since it can fit multiple TD types. Based on the selected severity levels and the selection, SonarQube detects 49 and 52 rules in Python and Java projects, respectively. The description of the rules and their classification can be found in Table~\ref{tab:RQ2} (shown in Appendix~\ref{append:rules}).

Based on the above, the variables that comprise the \textit{TD issues} for each unit of analysis are: 
(a) the count of self-fixed and non-self-fixed issues for each type of debt, calculated as the sum of the issues for all rules mapped to that type, 
(b) the count of self-fixed and non-self-fixed issues for each rule, and 
(c) the remediation time for each self-fixed and non-self-fixed issue.

\vspace{0.2cm}
\noindent
\textbf{Step 2: Snapshot Data Extraction}

For this step, we cloned the selected Apache Python and Java projects from GitHub. 

Then, we used a script to: 
(1) extract the entire change history of each project; 
(2) reserve at least the first 2,000 consecutive commits per project, therefore defining the snapshots (see Section \ref{subsec:Case Selection}); 
and (3) submit the snapshots to SonarQube in chronological order.
As mentioned before, for each snapshot, we recorded \textit{project name}, \textit{commit hashcode}, the changed files, and the \textit{developer} who is the author of the commit, including their \textit{name} and \textit{email}. Furthermore, we only analyzed the commits of the main branch of the studied systems, since the changes are inspected responsibly by the project's core team~\cite{kalliamvakou_2014MSR} and we focus on TD that has been repaid in the final product. We acknowledge that there are limitations with the use of such tools (e.g., thresholds associated with some detection rules), which are discussed in Section \ref{sec:Threats to Validity}.

\vspace{0.2cm}
\noindent
\textbf{Step 3: Self-Fixed Technical Debt Identification}

When analyzing multiple revisions of a system, SonarQube tracks the detailed information of issues that are fixed during the system evolution. 
For example, the first commit that contains the issue is considered as the commit that introduced it; similarly, the commit in which the issue disappeared is considered as the one that fixed that issue. 

According to the documentation of SonarQube\footnote{\url{https://docs.sonarqube.org/latest/user-guide/issues/}, visited in September 2020}, issues that are considered to be fixed could be involved in one of two situations: 
(1) issues have been corrected (i.e., the issues are truly fixed); 
or (2) the file is no longer available (removed from the project). 
The latter would be false positives in our study, as it is doubtful whether the developers actually aimed at fixing the technical debt by deleting a file.
To handle this problem, we filter out the issues that disappeared due to file deletion. 
The same method was used in some previous studies~\cite{DigkasSANER2018,Tan_JSME2020}.

To identify whether an issue is fixed by the same developer who introduces it, we search the \textit{snapshots} to find the \textit{developer} who introduced and fixed that issue. 
Furthermore, developers might submit a commit using a different email (e.g., work or personal) or version of the name (e.g., with acronyms). Thus, we compare developers' name and email address, considering it to be the same person when either the name or email address is the same.

\subsection{Data Analysis}
\label{sec:analysis_procedure}

To answer \textbf{RQ1}, we compare the number of TD issues that were self-fixed and non-self-fixed (i.e., fixed by another developer) for each project $p$. 
In particular, we calculate the \textbf{\textit{self-fixing\_rate}} per project, i.e., the percentage of TD issues that are fixed by the same developers in each project, which is measured as:
\begin{flalign} 
    \textnormal{\textit{self-fixing\_rate}}_{p} = \frac{\textnormal{count\_self-fixed}_{p}}{\textnormal{count\_fixed}_{p}} 
\end{flalign}
where $\textnormal{count\_self-fixed}_{p}$ is the number of self-fixed TD issues in project $p$ and $\textnormal{count\_fixed}_{p}$ is the total number of issues that have been fixed in the same project.

Furthermore, we build a generalized linear mixed model (GLMM)~\cite{GLMM_2005} to investigate the relationship between the likelihood of an issue being self-fixed and several project characteristics: the number of commits, number of developers, SLOC, and the total number of issues.

For \textbf{RQ2}, we calculate the \textbf{\textit{self-fixing\_rate}} per rule, i.e., the percentage of TD issues pertaining to rule $r$ that have been self-fixed in all projects:
\begin{flalign} 
    \textnormal{\textit{self-fixing\_rate}}_{r} = \frac{\sum_{p=1}^{n}\textnormal{count\_self-fixed}_{p,r} }{\sum_{p=1}^{n}\textnormal{count\_fixed}_{p,r}} 
\end{flalign}
where $n$ is the number of projects (i.e., 36), $\textnormal{count\_fixed}_{p,r}$ is the amount of issues that have been fixed for rule $r$ in project $p$ and $\textnormal{count\_self-fixed}_{p,r}$ is the number of issues for rule $r$ in project $p$ that are self-fixed. 

Subsequently, we compare the difference of the self-fixing rate for each debt type between Python and Java projects with Wilcoxon Rank Sum tests and calculate Cliff's Delta Effect Size. We also investigate which types of debt are more likely to be self-fixed and evaluate the significance of the observed differences by conducting a Scott-Knott Effect Size Difference test.

For \textbf{RQ3}, we calculate and analyze the survival time of each TD issue. 
This variable is measured as the number of days between the introduction of an issue and the moment when it is fixed in the source code. We use the Kaplan-Meier method to analyze and compare the survival time of the self-fixed and non-self-fixed issues.

To answer \textbf{RQ4}, we extract the detailed commit information to estimate how many developers maintain a given file $f$, i.e., $\textit{effort\_sharing}_{f}$. This value is extracted by calculating the amount of developers that changed a file $f$ from the time it was created to the last analyzed commit.

Using $\textit{effort\_sharing}_{f}$ alone to represent effort sharing may incur some bias. For example, we hypothesize that the fewer developers contribute to the file\footnote{By `file contribution' we mean that a commit contains a change (i.e., line addition, removal or edit) to a given file.}, the more likely they will modify their own code, implying a higher chance of fixing their own issues in that file. However, if these developers only make a few changes to the file, TD issues that were introduced by themselves may still remain in the project. Thus, investigating the number of changes that developers made to a file is also important. For that, we calculate an adjusted effort sharing using the following formula:

\begin{flalign} 
    \textnormal{\textit{effort\_sharing\_adj}}_{f} =  \frac{\sum_{d=1}^{\textnormal{count\_dev}_{f}}{\textnormal{change\_count}_{d,f}}}{\textnormal{count\_dev}_{f}}
\end{flalign}
where $\textnormal{count\_dev}_{f}$ is the amount of developers that changed a file $f$ from the time it was created to the last analyzed commit; $\textnormal{change\_count}_{d,f}$ is how many times a developer $d$ changed one file.
To assess the association between these two variables (i.e., $\textnormal{count\_dev}_{f}$ and $\textnormal{change\_count}_{d,f}$) and self-fixing, we build and examine a GLMM where the likelihood of an issue being self-fixed is predicted based on the two mentioned factors.

For \textbf{RQ5}, to investigate the association between independent variables related to developers and the likelihood of a developer to self-fix TD, we first introduce four variables.
\begin{itemize}
    \item The \textit{seniority} represents the number of days a developer has been involved in a project, and is based on Eyolfson et al. and Alfayez et al.~\cite{Eyolfson_2014,Alfayez_TechDebt18}. To calculate the seniority of each author, we first look for the author's last commit date in the analyzed commits and then subtract the author's first commit date. 
    \item The \textit{file count} refers to the number of files that a developer changed in throughout the analyzed history.
    \item The \textit{commit count} refers to the number of commits that a developer authored throughout the analyzed history.
    \item The \textit{commit size} refers to the average number of LOC that a developer contributed per commit throughout the analyzed history.
\end{itemize}
Similar to the previous research questions, 
we build a GLMM to assess the relationship between the developer-related variables and the likelihood of issues being self-fixed by them.

\section{Results}
\label{sec:Results}

\subsection{How much technical debt is self-fixed in open source projects?}

This study required the mining of over 17K commits from 20 Python projects and 27K commits from 16 Java projects. After filtering out issues that were fixed due to file deletion (see Section \ref{sec:variables}), 54K Python issues and 71K Java issues have been selected in total, among which 35K and 34K, respectively, were fixed.

\begin{table}[H]
	\caption{Details of the 20 Python and 16 Java Projects}
	\footnotesize
	\setlength{\tabcolsep}{4pt}
	\begin{center}
		\begin{tabular}{lrrrrrrrr}
			\toprule
			& \multicolumn{3}{c}{\textbf{Python}} & \multicolumn{3}{c}{\textbf{Java}} & \multicolumn{2}{c}{\textbf{All}}\\
			\cmidrule(r){2-4}\cmidrule(r){5-7}\cmidrule(r){8-9}
			& \textbf{Median} & \textbf{Average} & \textbf{Max} & \textbf{Median} & \textbf{Average} & \textbf{Max} & \textbf{Median} & \textbf{Average} \\
			
			\midrule
			\textbf{Commits} & 337 & 853 & 2443 & 1454 & 1760 & 3799 & 1222 & 1227\\
			\textbf{SLOC} & 5307 & 14983 & 74658 & 87637 & 110881 & 316552 & 39043 & 56242\\
		    \textbf{Developers} & 9 & 18 & 85 & 55 & 89 & 319 & 21 & 48\\
			\textbf{Total Issues} & 542 & 2679 & 24931 & 3518 & 4994 & 9904 & 2686 & 3723\\
			\textbf{Fixing Rate} & 48.85\% & 51.91\% & 86.34\% & 39.78\% & 44.44\% & 81.08\% & 48.86\% & 49.06\%\\
			\textbf{Self-fixing Rate} & 67.21\% & 65.54\% & 100.00\% & 28.99\% & 33.26\% & 93.86\% & 47.64\% & 50.84\%\\
			\bottomrule
		\end{tabular}
		\label{tab:RQ1}
	\end{center}
\end{table}

Table~\ref{tab:RQ1} provides the following details about the 36 selected projects: 
the number of commits, SLOC, the number of developers involved in our study (i.e., in the analyzed commits), the absolute number of total issues, fixing rate, and the rate of the self-fixed issues.
As shown in Table~\ref{tab:RQ1}, Python projects seem to have a higher fixing rate and self-fixing rate than Java projects in general. Especially, about two thirds of the fixed Python issues are fixed by the same developers who introduced them, while this proportion is only one third in Java projects. Considering all projects together (Python and Java), almost half of the issues have been fixed, and among them, about half of the fixed issues were self-fixed.
The results indicate that \finding{about half of the technical debt seems to be paid back by the same developers who introduced them.}

From Table~\ref{tab:RQ1}, we also observe that the projects have different characteristics, i.e., the number of commits, developers, SLOC, the number of total issues, which might affect the self-fixing rates of different projects. To estimate how these characteristics relate to self-fixing, we build a generalized linear mixed model (GLMM)~\cite{GLMM_2005}, where the dependent variable captures, for each TD issue, whether it is self-fixed or non-self-fixed. The random effect is represented by the project ID, and helps to account for differences between projects in terms of the mentioned characteristics. In addition, to properly interpret the importance of each variable in the model, we use min-max normalization to normalize variable values, within each characteristic, in the interval [0,1]\footnote{This is done by subtracting the minimum and dividing by the difference between the maximum and minimum, similarly as was performed in~\cite{iammarino_2021JSS_SATD_model}}.
Moreover, we use the whole dataset from Python and Java projects to perform the analysis because we intend to build an explanatory model, not a predictive model.

To avoid multi-collinearity, we use the R function \emph{redun}, of the \emph{Hmisc} package~\cite{hmisc_2019}, for removing redundant variables. The function \emph{redun} removes variables stepwise, starting from the most predictable one until no variable can be predicted with an adjusted $R^2$ greater than a given threshold, i.e., 0.8 in this study. The result of executing \emph{redun} shows that there is no redundant variable. Then, we build the generalized linear mixed model using the function \emph{glmer} of the \emph{lme4} R package~\cite{lme4_2007}. Table~\ref{tab:RQ1_model} reports the results of the mixed-effect regression model. The top part of the table reports the model diagnostics, i.e., Akaike Information Criterion (AIC), Bayesian Information Criterion (BIC), log likelihood, deviance and degree of freedom residuals, the scaled residuals and the random effects. The bottom part of the table reports the odds ratio (OR), estimate, standard error, z-value and $p$-value\footnote{Among these values, the $p$-value indicates whether the factor is statistically significant or not (for a significance level of 95\%)}.

As shown in the table, SLOC, the number of developers and the total number of TD issues have a statistically significant effect on the likelihood of an issue being self-fixed. The result indicates that \finding{the size of the project, the number of developers involved in it, and the overall quality of the project (i.e., the total number of TD issues in this study) are associated with whether TD items are self-fixed, while the number of commits has no significant relation.}

In addition, 
Table~\ref{tab:RQ1_model} shows that the increase of SLOC, number of developers and number of total issues can decrease the likelihood that the issue will be self-fixed (i.e., OR \textless 1)\footnote{OR is given by $e^{c_i}$ where $c_i$ is the coefficient of the $i$th factor for a logistic regression model. An OR \textgreater 1 indicates that an increase of variable increases of OR times the chances of an issue to be self-fixed.}. The results suggest that \finding{as a project evolves, growing in size, number of developers and total issues, the likelihood of its TD issues to be self-fixed tends to decline}.

\begin{table}[H]
	\caption{Results of the generalized linear mixed model}
	\small
	\begin{center}
		\begin{tabular}{lrrrrr}
			\toprule
            Diagnostics \\
            \midrule
            & AIC & BIC & logLik & deviance & df resid. \\
			& 74950.6 & 75005.5 & -37469.3 & 74938.6 & 68945 \\
            \midrule
            Scaled residuals \\
            \midrule
            & Min & 1Q & Median & 3Q & Max \\
            & -11.338 & -0.591 & -0.510 & 0.912 & 4.368 \\
            \midrule
            Random effects\\
			\midrule
			Groups name & Variance & Std.Dev. \\
            Project ID (Intercept) & 3.944 & 1.986 \\
            \midrule
            Quality metrics \\
            \midrule
            Metric & OR & Estimate & Std.Error & z-value & $p$-value \\
            \midrule
            (Intercept) & 5.085 & 1.626 & 0.320 & 5.082 & \textless \textbf{0.001} \\
            SLOC        & 0.115 & -2.164 & 0.600 & -3.606 & \textless \textbf{0.001} \\
            Commits     & 1.318 & 0.276 & 0.483 & 0.573 & 0.567 \\
            Developers  & 0.138 & -1.981 & 0.485 & -4.082 & \textless \textbf{0.001} \\
            Total Issues & 0.033 & -3.426 & 0.501 & -6.844 & \textless \textbf{0.001} \\
			\bottomrule
		\end{tabular}
		\label{tab:RQ1_model}
	\end{center}
\end{table}

\subsection{Which type of technical debt is more likely to be self-fixed?}
\label{sec:RQ2}

\begin{figure}
	\centering 
	\includegraphics[width=1.2\columnwidth]{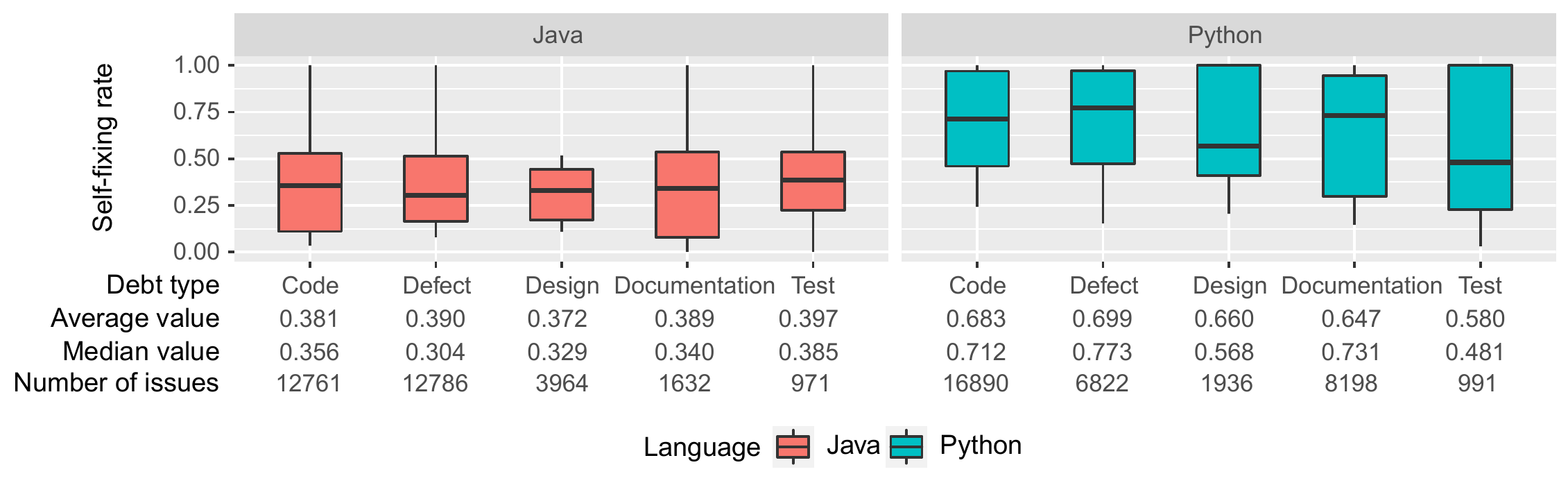}
	\caption{Distribution of self-fixing rates of five TD types among 20 Python and 16 Java projects}
	\label{fig:rq2} 
\end{figure}

Figure~\ref{fig:rq2} shows a series of box plots depicting the distributions of self-fixing rates of the five studied TD types among the 20 Python and 16 Java projects, together with the average and median values, and the number of fixed issues (bottom of figure). According to Figure~\ref{fig:rq2}, the self-fixing rate for each debt type in Python projects is nearly double compared to Java projects. To further evaluate the significance of the difference for each type of debt, we calculated the Wilcoxon Rank Sum test~\cite{Wilcoxon_2014} and Cliff's Delta Effect Size~\cite{cliff_2014} on the self-fixing rates of the Java and Python projects, since we cannot assume the rates are normally distributed. Table~\ref{tab:RQ2_wilcoxon} shows the results of the statistical tests, which reveal that except for Test Debt, the difference between the self-fixing rates of each type in Python and Java projects is significant (i.e., $p$-value \textless 0.05) and with a medium effect size\footnote{The magnitude is assessed using the thresholds provided by Romano et al.~\cite{delta_2006}, i.e. $|delta|$ \textless 0.147 is negligible, $|delta|$ \textless 0.33 is small, $|delta|$ \textless 0.474 is medium, otherwise is large} at least. Thus, issues related to most types of TD are more likely to be self-fixed in Python projects than in Java projects.

\begin{table}[H]
	\caption{Results of Wilcoxon Rank Sum tests and Cliff's Delta Effect Size}
	\small
	\begin{center}
		\begin{tabular}{lrrrrr}
			\toprule
            & Code & Defect & Design & Documentation & Test\\
            \midrule
            $p$-value & 0.006**  & 0.006** & 0.005** & 0.034* & 0.253\\
			$delta$ & -0.541 (large) & -0.544 (large) & -0.550 (large) & -0.427 (medium)& -0.252 \\
			\bottomrule
		\end{tabular}
		\label{tab:RQ2_wilcoxon}
	\end{center}
\end{table}

To investigate the self-fixing rate of TD in more depth, we consider the numerous rules for the different TD types (see Section~\ref{sec:variables}). 
Table~\ref{tab:RQ2} (in Appendix~\ref{append:rules}) shows for each rule:
the ID number, type and definition of the rule, 
self-fixing rate, issue fixing rate, the total number of issues and the number of projects in which violations of the rule appear.
In total, Table~\ref{tab:RQ2} includes 49 Python and 52 Java rules, and they are sorted in decreasing order of self-fixing rates.

Table~\ref{tab:RQ2} includes 38, 41, 14, 4 and 4 rules related to Code Debt, Defect Debt, Design Debt, Documentation Debt and Test Debt, respectively. The self-fixing rates vary widely between different types of debt, i.e., from 5\% to 100\%. To investigate which types of debt are more likely to be self-fixed and evaluate the significance of the self-fixing rates among different types, we conducted a Scott-Knott Effect Size Difference (ESD) test~\cite{tantithamthavorn_2016TSE_ESD} to group the five TD types into statistically distinct ranks based on the their self-fixing rates.
As a variant of the Scott-Knott test~\cite{scott_1974}, the Scott-Knott ESD test evaluates a non-normally distributed dataset and merges any two statistically distinct groups that have a negligible effect size into a single group~\cite{tantithamthavorn_2016TSE_ESD}. Table~\ref{tab:RQ2_Scott} shows the rank of the five TD types based on their self-fixing rates. We found that the five TD types are distributed in three distinct groups and, thus, the self-fixing rates are significantly different between groups. The self-fixing rates of Defect Debt and Code Debt (group \#1) are considerably higher than the other three; in other words, \finding{Defect Debt and Code Debt receive more attention from developers who introduced them in the projects}. In contrast, Test Debt was ranked lowest (group \#3) according to their self-fixing rates, indicating that \finding{Test Debt is less likely to be self-fixed}.

\begin{table}[H]
	\caption{Ranks of five debt types according to the Scott-Knott ESD tests}
	\small
	\begin{center}
		\begin{tabular}{cl}
			\toprule
            Group & Types of TD\\
            \midrule
            1 & Defect Debt, Code Debt\\
            2 & Design Debt, Documentation Debt\\
            3 & Test Debt\\
			\bottomrule
		\end{tabular}
		\label{tab:RQ2_Scott}
	\end{center}
\end{table}

\subsection{How long does self-fixed technical debt survive during the evolution of a system?}

\begin{figure}
	\centering 
	\includegraphics[width=1\columnwidth]{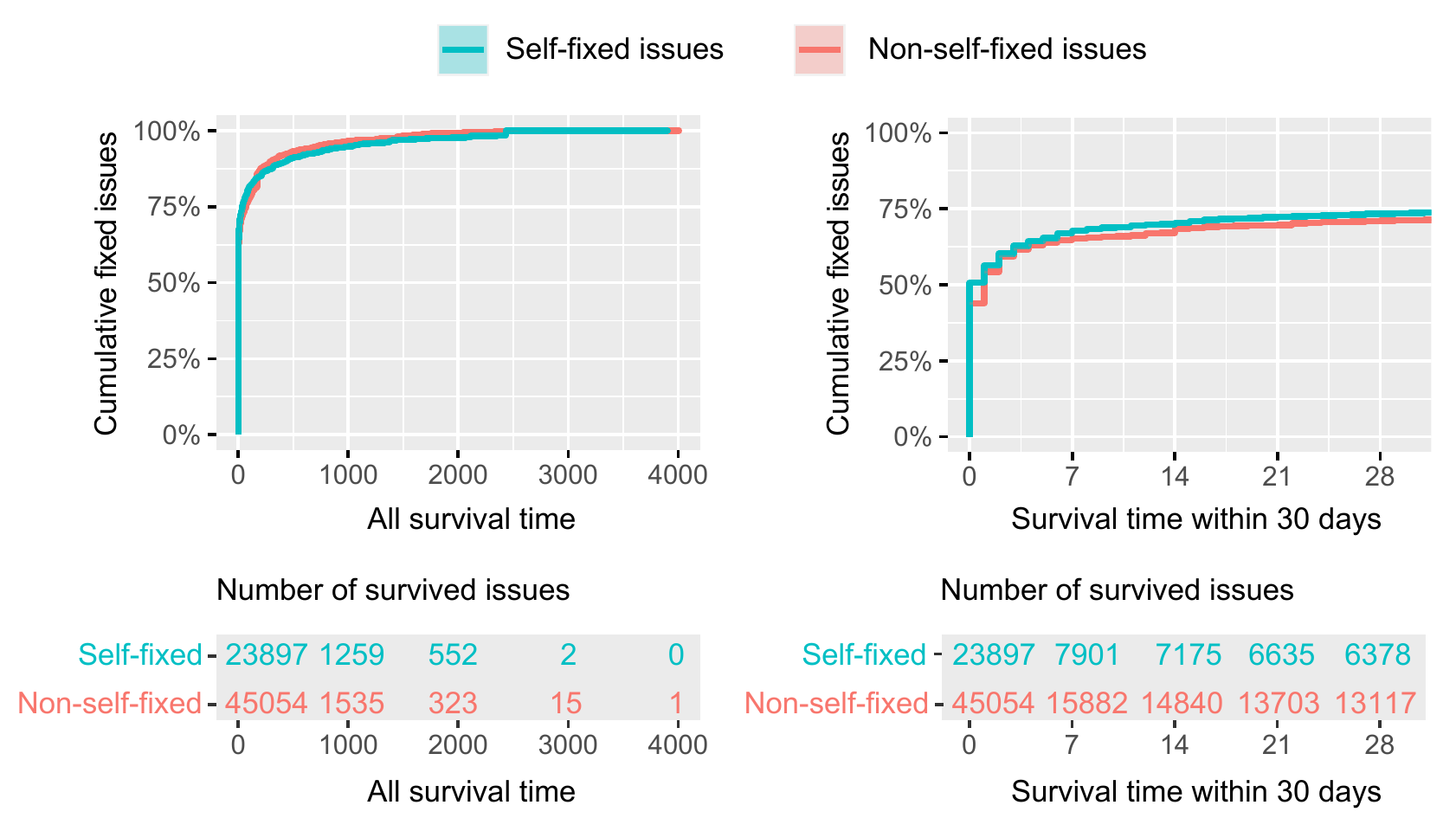}
	\caption{Results of Kaplan-Meier method for the survival time of self-fixed and non-self-fixed issues}
	\label{fig:RQ3_all} 
\end{figure}

To answer this research question, we used the Kaplan–Meier (K–M) method to analyze the survival time of the fixed issues, which is a non-parametric statistic~\cite{kaplan_1958}. Figure~\ref{fig:RQ3_all} shows the Kaplan-Meier plots to visualize survival curves for all the self-fixed and non-self-fixed issues. In Figure~\ref{fig:RQ3_all}, the left figure shows all fixed issues and their entire life cycle, i.e., the number of days between introduction and repayment. The y-axis represents the cumulative percentage of TD issues that have been fixed within a specified number of days (x-axis). In addition, the number of surviving issues is shown at the bottom of the figure. For example, there are 23,897 self-fixed issues and 45,054 non-self-fixed issues in the beginning of the projects, while 1,259 and 1,535 issues survived for more than 1,000 days, respectively.

From the left side of Figure~\ref{fig:RQ3_all}, we found that most of both self-fixed and non-self-fixed issues were paid back in a short period of time. Since most issues have been fixed quickly, we zoomed in on the issues fixed within one month as shown on the right side of Figure~\ref{fig:RQ3_all}. As shown in the figure, the survival curves of self-fixed and non-self-fixed issues almost overlap, and the percentage of self-fixed issues is pretty similar to that of the non-self-fixed issues.

\begin{figure}
	\centering 
	\includegraphics[width=0.8\columnwidth]{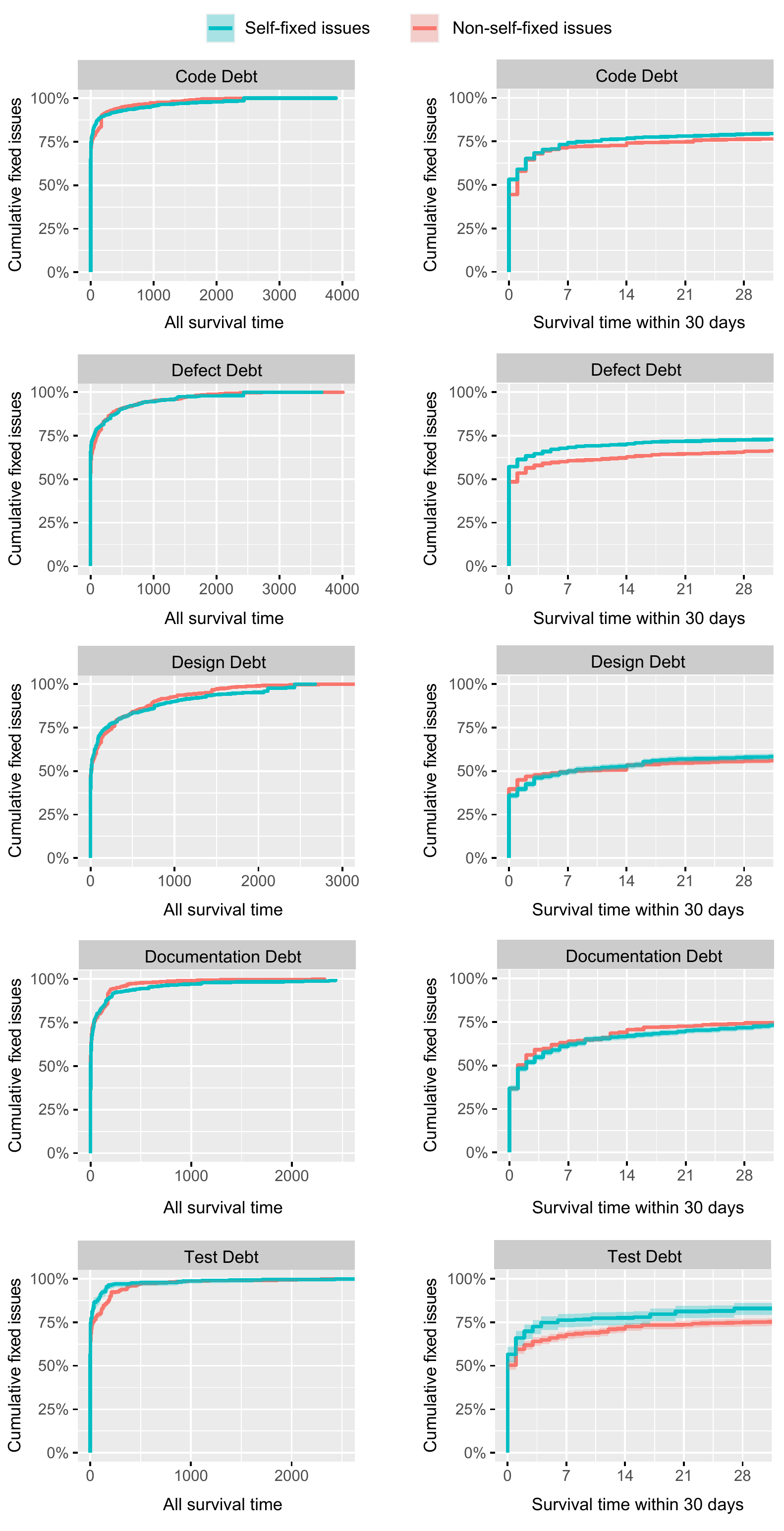}
	\caption{Results of Kaplan-Meier method for the survival time for self-fixed and non-self-fixed issues belonging to different types}
	\label{fig:RQ3_types} 
\end{figure}

To further investigate the survival time of different types of self-fixed issues and compare them with the non-self-fixed issues, we calculated the survival time for all issues belonging to the different types. The results are shown in Figure~\ref{fig:RQ3_types}. Similar to Figure~\ref{fig:RQ3_all}, the left side depicts all fixed issues in their entire life cycle for each TD type, and the right side zooms in the issues fixed within one month. A visual inspection shows that there are some differences between self-fixed and non-self-fixed issues in the 30-day window, but they do not appear to be major.

Table~\ref{tab:RQ3_percentage} shows how often an issue was fixed within a certain time-frame (i.e., one week, one month and one year) for each type of debt. 
To investigate whether the self-fixed issues are fixed more quickly than non-self-fixed issues, we conducted pairwise Fisher's exact tests~\cite{fisher_1922} on the percentages of self-fixed and non-self-fixed issues.
The results confirm the aforementioned visual inspection, showing that, \finding{for individual types of TD, there is no significant difference between the percentages of self-fixed and non-self-fixed issues within each time frame ($p$-values \textgreater 0.05).}

\begin{table}
	\caption{Percentages of self-fixed and non-self-fixed issues within different time periods}
	\footnotesize
	\begin{center}
		\begin{tabular}{llrrrrr}
			\toprule
			& & \textbf{Code} & \textbf{Defect} & \textbf{Design} & \textbf{Documentation} & \textbf{Test} \\
			\midrule
			\multirow{2}{*}{\textbf{$\leq$ 1 week}} & Self-fixed & 73.25\% & 67.80\% & 49.09\% & 60.08\% & 76.25\% \\
			 & Non-self-fixed & 71.41\% & 60.01\% & 49.34\% & 62.99\% & 67.00\%\\
			\multirow{2}{*}{\textbf{$\leq$ 1 month}} & Self-fixed & 79.33\% & 72.69\% & 58.00\% & 72.34\% & 82.92\% \\
			 & Non-self-fixed & 76.20\% & 66.06\% & 55.74\% & 74.37\% & 75.03\% \\
			\multirow{2}{*}{\textbf{$\leq$ 1 year}} & Self-fixed & 91.78\% & 86.98\% & 80.43\% & 93.28\% & 97.08\% \\
			 & Non-self-fixed & 93.54\% & 88.02\% & 80.43\% & 96.27\% & 93.93\% \\
			\bottomrule
		\end{tabular}
		\label{tab:RQ3_percentage}
	\end{center}
\end{table}

Focusing on the results related to self-fixing in Table~\ref{tab:RQ3_percentage}, it seems that the issues related to Code Debt, Defect Debt and Test Debt are repaid in a shorter time. To further investigate whether the issues belonging to one TD type are self-fixed faster or slower than other types, we conducted a Scott-Knott Effect Size Difference (ESD) test~\cite{tantithamthavorn_2016TSE_ESD} to group the five TD types into statistically distinct ranks based on their percentages for the three time-frames. Table~\ref{tab:RQ3_Scott} shows the rank of the five TD types based on their percentages of issues that have been self-fixed within the three time-frames.  We found that the five TD types are distributed in four distinct groups and, thus, the percentages of self-fixed issues for all time-frames are significantly different between groups. Since the percentages of Test Debt (group \#1) and Code Debt (group \#2) are considerably higher than the others, it implies that \finding{Test Debt tends to be self-fixed faster than the other four debt types, followed by Code Debt}. On the contrary, Design Debt was ranked lowest (group \#4) according to their percentages, which indicates that \finding{self-fixed issues related to Design Debt have a longer survival time}.

\begin{table}
	\caption{Ranks of five debt types according to the Scott-Knott ESD tests}
	\small
	\begin{center}
		\begin{tabular}{cl}
			\toprule
            Group & Types of TD\\
            \midrule
            1 & Test Debt\\
            2 & Code Debt\\
            3 & Defect Debt, Documentation Debt \\
            4 & Design Debt\\
			\bottomrule
		\end{tabular}
		\label{tab:RQ3_Scott}
	\end{center}
\end{table}

\subsection{How does the number of developers and the frequency of their modifications relate with the likelihood of TD issues to be self-fixed?}
\label{sec:RQ4}

This research question concerns effort sharing (i.e. the number of developers that maintain a given file) and adjusted effort sharing (i.e., the number of changes that developers made to a file). To examine these factors, we build a generalized linear mixed model~\cite{GLMM_2005} to estimate whether and how an issue that has been self-fixed correlates with the effort sharing and adjusted effort sharing of the corresponding file in which the issue was detected. In the model, the random effect is represented by the project and file ID, and helps to account for differences between different files in terms of effort sharing and adjusted effort sharing. Moreover, we use the whole dataset (both Python and Java projects) to perform the analysis because we intend to build an explanatory model, not a predictive model. In addition, the value of the independent variables can depend on projects' characteristics, e.g., the larger a project is, the more developers might be involved in it. To properly interpret the importance of each variable in the model, we normalize variable values, within each project, in the interval [0,1]\footnote{This is done by subtracting the minimum and dividing by the difference between the maximum and minimum, which was also used in~\cite{iammarino_2021JSS_SATD_model}.}. After that, similarly to RQ1, we build the generalized linear mixed model using the \emph{glmer} function of the \emph{lme4} R package~\cite{lme4_2007}.

Table~\ref{tab:RQ4_model} reports the results of the logistic regression mixed-effect model; it can be read similarly to Table~\ref{tab:RQ1_model}. 
The results show that both effort sharing and adjust effort sharing have a statistically significant effect on the likelihood of an issue being self-fixed; in other words, \finding{the number of developers involved in the files and the number of changes that developers make per file can affect the probability of the contained issues being self-fixed.}

\begin{table}[H]
	\caption{Results of the generalized linear mixed model}
	\small
	\begin{center}
		\begin{tabular}{lrrrrr}
			\toprule
            Diagnostics \\
            \midrule
            & AIC & BIC & logLik & deviance & df resid. \\
			& 56735.7 & 56781.1 & -28362.8 & 56725.7 & 65529\\ 
            \midrule
            Scaled residuals \\
            \midrule
            & Min & 1Q & Median & 3Q & Max \\
            & -8.5715 & -0.5055 & -0.1879 & 0.2854 & 9.4512  \\ 
            \midrule
            Random effects\\
			\midrule
			Groups name & Variance & Std.Dev. \\
            File ID (Intercept) & 5.387 & 2.321 \\
            Project ID (Intercept) & 11.298 & 3.361 \\
            \midrule
            Quality metrics \\
            \midrule
            Metric & OR & Estimate & Std.Error & z-value & $p$-value \\
            \midrule
            (Intercept) & 2.773 & 1.020 & 0.472 & 2.160 & 0.03 \\
            effort sharing & 0.064 & -2.755 & 0.279 & -9.870 & \textless \textbf{0.001} \\
            adjust effort sharing & 23.492 & 3.157 & 0.362 & 8.714 & \textless \textbf{0.001} \\
			\bottomrule
		\end{tabular}
		\label{tab:RQ4_model}
	\end{center}
\end{table}

In addition, the increase of adjusted effort sharing (i.e., based on the number of changes) can increase by about 23.5 times the odds of an issue being self-fixed. However, the increase of effort sharing (i.e., based on the number of developers) can decrease the likelihood of the issue to be self-fixed (OR \textless 1). The results indicate that \finding{the fewer developers maintaining a file and the more changes made per file, the higher the chance of an issue being self-fixed.} More generally, we argue that when investigating effort sharing, it is not enough to look at the number of developers that maintain a given file, but one should also consider the amount of changes per developer.

\subsection{Who (self) fixes technical debt?}
\label{sec:RQ5}

The study involved 180 Python developers and 331 Java developers that have fixed technical debt during the evolution of the 20 Python and 16 Java projects. Among them, 96 Python and 141 Java developers have self-fixed TD. To answer this research question, we investigate the relationship between self-fixed issues and developers' \textit{seniority}, \textit{file count}, \textit{commit count} and \textit{commit size}.

The seniority refers to how long a developer has contributed to a project, calculated as the number of days between the developer's last and first commit date (see Section~\ref{sec:analysis_procedure}). However, if a developer has been working on a project for a long time but has contributed few commits or to few files, then this developer tends to be unfamiliar with the progress update and feel less responsible for the project. 
Conversely, experienced developers might be over-confident in modifying unfamiliar code, and thus inadvertently lower its quality. Therefore, it is paramount to also examine variables that reflect a developer's level of involvement in the project, i.e., the commit count and file count. We assess the two variables based on the commits performed by a developer (see Section~\ref{sec:analysis_procedure}). Moreover, some developers may commit fewer times or participate in fewer files, but they contribute more to some key components, which may also increase the likelihood that they will repay their own debt. Thus, we also consider the commit size, i.e., the average number of LOC that a developer contributed per commit (see Section~\ref{sec:analysis_procedure}).

To investigate the relationship between the mentioned variables and self-fixing, we build a GLMM where the dependent variable captures, for each TD issue, whether it is self-fixed or non-self-fixed. The random effect is represented by the project and developer ID, and helps to account for differences between developers in terms of seniority, file count, commit count and commit size. To properly interpret the importance of each variable in the model, we normalize variable values, within each project, in the interval [0,1]\footnote{This is done by subtracting the minimum and dividing by the difference between the maximum and minimum, which was also used in~\cite{iammarino_2021JSS_SATD_model}.}

Table~\ref{tab:RQ5_model} reports the results of the logistic regression mixed-effect model and is interpreted as Table~\ref{tab:RQ1_model}.
From the table, we observe that seniority and commit count have a statistically significant effect on the likelihood of an issue being self-fixed. The result indicates that \finding{the number of days developers have been involved in a project (i.e., seniority) and the number of commits they authored (i.e., commit count) relate to whether they self-fix TD. In contrast,  the number of files to which they have contributed (i.e., file count) and the average number of LOC that developers contribute per commit (i.e., commit size) have no significant effect.}

\begin{table}[H]
	\caption{Results of the generalized linear mixed model}
	\small
	\begin{center}
		\begin{tabular}{lrrrrr}
			\toprule
            Diagnostics \\
            \midrule
            & AIC & BIC & logLik & deviance & df resid. \\
			& 56113.0 & 56176.2 & -28049.5 & 56099.0 & 61836 \\
            \midrule
            Scaled residuals \\
            \midrule
            & Min & 1Q & Median & 3Q & Max \\
            & -8.9266 & -0.6001 & -0.1930 & 0.6890 & 15.9939 \\
            \midrule
            Random effects\\
			\midrule
			Groups name & Variance & Std.Dev. \\
            Developer ID (Intercept) & 2.800 & 1.673 \\
            Project ID (Intercept) & 2.713  & 1.647 \\
            \midrule
            Quality metrics \\
            \midrule
            Metric & OR & Estimate & Std.Error & z-value & $p$-value \\
            \midrule
            (Intercept) & 0.513 & -0.667 & 0.406 & -1.643 & 0.100 \\
            Seniority & 0.380 & -0.968 & 0.345 & -2.807 & \textbf{0.005} \\
            File count & 1.260 & 0.231 & 0.524 & 0.441 & 0.659 \\
            Commit count & 7.849 & 2.060 & 0.406 & 5.071 & \textless \textbf{0.001} \\
            Commit size & 1.163 & 0.151 & 0.343 & 0.440 & 0.660 \\
			\bottomrule
		\end{tabular}
		\label{tab:RQ5_model}
	\end{center}
\end{table}

In addition, the increase of commit count can increase by about 7.8 times the odds of an issue being self-fixed. However, the increase of seniority can decrease this likelihood (OR \textless 1). These results indicate that \finding{developers who contributed more commits are more likely to fix their debt. Furthermore, the longer the developers are involved in a project, the less often they self-fix TD.}

\section{Discussion}
\label{sec:Discussion}

In this section, we first discuss the interpretation of the findings presented in Section~\ref{sec:Results}. Then, we describe the implications for researchers and practitioners.

\subsection{Interpretation of Results}

\subsubsection{The Likelihood of TD to be Self-fixed}

The phenomenon of self-fixed TD, as examined in this study, is widely spread in the analyzed Python and Java projects, as we observed that around half of fixed issues are self-fixed.
This indicates that these developers tend to clean up their own work, paying back the technical debt they incurred themselves. 
However, we also found that the likelihood of TD issues to be self-fixed is negatively associated with the number of developers, the total number of issues and SLOC of projects. 
Thus, our results suggest that, as a project grows in size, number of developers and total issues, one can expect a decline in self-fixing rate.

To expand our understanding of the phenomenon, we delved into certain human factors related to how developers maintain the projects. In particular, we investigated the distribution of effort among components from two perspectives. First, we looked from the point of view of teams and found that, when multiple developers work in a same file, the likelihood of self-fixed TD issues contained in the file and the overall project tends to decline as teams get larger. Second, we looked from the point of view of individuals and found that developers who contributed more commits are more likely to self-fix TD, while a longer involvement in the project reduces the chances of self-fixing TD. This may be an indication that a greater dedication (e.g., by providing more commits) but not seniority may reflect on a higher chance of self-fixing TD. Although we expected that commit size would render interesting observations, it displayed no significant effect; exploring the reason why is an interesting path for further research.

Putting the presented findings together, we note that even if fostering dedication among developers sounds like a good strategy, it is not necessarily desirable to push all developers to be extra active, as it would naturally lead to higher levels of collaboration on the same files, which would in turn drive the overall likelihood of TD being self-fixed down.
We also note that merely driving the likelihood of TD being self-fixed up should not be a strategy, as some development strategies that serve various purposes (incl. TD remediation) can also result in lower likelihood of TD being self-fixed. Examples of such strategies include rotation of developers, contributions from external developers (e.g., via pull requests) and maintenance tactics (e.g., assigning certain types of TD items to newcomers). 

\subsection{Differences between Python and Java}

The results of RQ2 (see Section~\ref{sec:RQ2}) reveal that, except for Test Debt, the difference between the self-fixing rates of each type in Python and Java projects is significant. Especially for Code Debt, Defect Debt and Design Debt, there are large effect sizes of the differences between the self-fixing rates of the two languages. 

One possible reason may be related to Python's dynamic type system, which makes it popular for flexibility, expressiveness and succinctness, but also less maintainable and secure~\cite{dyamically_typed_languages}. Developers may spend extra effort on software maintenance~\cite{PythonFeatureSEKE} and software quality improvement~\cite{PythonQuality} (related to Code Debt and Design Debt), especially since Python code is more change-prone due to having a higher number of dynamic features~\cite{PythonFeatureSEKE}. In addition, the misuse of dynamic features could lead to coding issues that are often fixed by adding exception handling (Defect Debt remediation)~\cite{PythonFeatureChina}. Altogether, Python developers may be more concerned about the issues they introduce themselves, making such TD more likely to be fixed in Python projects than in Java projects.

\subsubsection{Types of Technical Debt and Survival Time}

It is worth noting that Defect Debt and Code Debt are more likely to be self-fixed than other debt types. 
This may be an indication that developers are more aware of these two types or that it is easier to fix those issues\footnote{This observation may also be related to the definition of those types in SonarQube (see Section~\ref{sec:Threats to Validity}).}. However, the Defect Debt issues seem to survive a bit longer than Code Debt issues before being self-fixed (according to the grouping results of a Scott-Knott ESD test).  
These observations suggest that although Defect Debt and Code Debt are preferably self-fixed and by the majority of developers (i.e., regardless of the project and team setting), developers may prioritize Code Debt.

Design Debt has quite a low self-fixing rate and this is worrisome. 
One study by Besker et al.~\cite{Bill_ICSME2017} shows that TD related to design has the greatest negative impact on the daily software development activities. 
This entails that Design Debt can be risky and that the person that may know most about the code (i.e., the developer who incurred the debt) is not addressing it sufficiently. Moreover, the situation seems worse in larger projects, where the self-fixing rate of Design Debt issues is even lower. To verify that, we calculated the Spearman correlation between the size of the project (i.e., SLOC) and self-fixing rate of Design Debt, which showed to be strongly negative\footnote{We interpret the correlation coefficient according to Cohen~\cite{Spearman}, i.e., a strong correlation is when $0.5 \leq |\rho| \leq 1$} ($\rho = -0.58$). 
Furthermore, we found Design Debt issues to have the longest survival time among five debt types.
One way to tackle problems related to growing Design Debt, might be to focus on developers with expertise on this type of debt rather than encouraging self-fixing.

Finally, we found that developers are least likely to self-fixed Test Debt. However, when developers decided to pay back their Test Debt, it was done in the shortest possible time. These results need further investigation as only four rules related to Test Debt were considered in our study (i.e. they were detected by SonarQube); 
Other related studies analyzed ten or more test smells, some of which can lead to accumulating Test Debt~\cite{bavota2015test,Palomba:2016:Test}.

In summary, the results could be an indicative that developers are keen to resolve their own issues when they are related to decreasing code quality (Code Debt) or increasing chance of bugs and uncorrected known defects (Defect Debt). 
In contrast, developers may pay less attention to their own issues when they are related to Design Debt (e.g., long files and spaghetti code), which can affect understandability and maintainability and happen to be more often fixed by others.

\subsubsection{Human Factors}
\label{sec:DiscussionHF}

We found that the more developers are maintaining a file, the lower the chance of an issue being self-fixed. However, the number of files developers contribute to does not appear to impact the likelihood of issues being self-fixed. Thus, strategies such as encouraging developers to take over more components (i.e., contribute to more files) may not boost TD remediation.

Furthermore, we found that developers who contribute more commits are more likely to fix their TD. This may be because these developers are more experienced or have a better understanding of the project code. Such an explanation would align with the findings by Alfayez et al.~\cite{Alfayez_TechDebt18} that developers with higher experience in the projects are less likely to introduce technical debt.

In a somewhat contrasting finding, Potdar et al.~\cite{SATDICSME2014} observed that developers with more experience tend to introduce more SATD. However, such a difference may be due to the nature of SATD, i.e., developers admitting debt. The more experienced developers are, the easier it is to identify TD in the source code and document it. We clarify that the results of Potdar et al. do not contradict ours since they discuss the introduction of TD (and not repayment). Furthermore, other studies showed that the experience of developers fixing SATD is higher than that of developers introducing SATD~\cite{SATDMSR2016}, and that the majority of SATD is self-fixed~\cite{RemovalICSME2017}. 

\subsection{Implications for Researchers and Practitioners}

The findings and discussion drawn from this study can benefit both researchers and practitioners. 

\subsubsection{Implications for Researchers}

Researchers can use the self-fixing rates to develop tools to prioritize TD remediation by assigning different weights to TD issues and giving suggestions to developers. Such suggestions could concern what types of issues to repay faster and whether they should pay the issues themselves or not. For example, TD tools could highlight critical issues that tend to survive for a longer period of time. The developer who incurred it could be notified first and, if she does not fix it, other team members could be involved. 

Such tools could also monitor the source code and warn developers when an entire component is becoming a hot-spot of (potentially) long-living issues. The involved developers could be then notified and decide whether the involved TD issues should be fixed, delegated to other developers, or even ignored. Moreover, investigating how development behaviors (e.g., effort sharing) and developers' characteristics (e.g., seniority) are associated with the likelihood of the TD contained in a project being self-fixed may provide insights for predicting which projects (or components) are more likely to have unpaid TD. For example, if there is a lack of experienced developers contributing to a component, team leaders could ask more experienced developers to assess the severity of the accumulated debt. If necessary, these developers could also be allocated to test and maintain the component. 

Furthermore, researchers can use our findings to design different software maintenance strategies tailored for a variety of team configurations. For example, for parts of a project that require urgent delivery, the team leader can be suggested to coordinate with experienced developers to focus on those parts. Because the experienced developers who concentrate on a certain component are better at self-fixing their TD, this practice may boost the quality of the code that has to be written in the short term (TD will be fixed quicker and better).

\subsubsection{Implications for Practitioners}

Practitioners can use the results to guide their refactoring and maintenance activities. Specifically, software development teams can use them to better assign tasks among different developers and adjust team collaboration. For example, Code Debt and Defect Debt that need to be addressed urgently can be assigned to the developer who introduced them; in contrast, Design Debt can better be assigned to other developers because it is not often self-fixed, and tends to survive longer.
By knowing which TD types are more likely to be fixed in the near future,
both developers and team leaders can concentrate on TD that is already overdue and has not been addressed, and on who is most suited to fix it. 

Our investigation on how human factors are associated with the phenomenon of self-fixed TD can directly support practitioners to re-allocate suitable developers to components. For example, consider a component that displays a high incidence of TD and is maintained by, among others, self-fixing-prone developers. If the accumulation of TD in the component is related to time pressure (i.e., the developers have knowledge to repay it but not enough time), a good strategy may be to let them focus their effort further on this component by reducing their workload on other components; this could effectively reduce TD through intensive self-fixing. In addition, other developers  (with lower self-fixing rates) may be re-allocated to different components; this may give them an opportunity to improve their skills and knowledge of the project and become themselves self-fixing-prone developers.

Moreover, knowledge of human factors may further inform which components are more TD-prone (or less fix-prone) so that more testing effort can be budgeted for them. When maintaining those components, developers could also be encouraged to address their TD issues that belong to the debt types with a high self-fixing rate, since those issues might be more familiar and easier for them.

Finally, teams that already manage their TD by other means, can compare their current practice with our results or the results from replicating our study in their own source code, allowing for more informed decisions. For example, if a team has several open issues related to software comprehension and maintenance (e.g., long file and spaghetti code), they should be aware that the developers who introduced them are not likely to fix the issues in a short time.

\section{Threats to Validity}
\label{sec:Threats to Validity}

In the following, we discuss the threats to construct and external validity of the reported study, as well as reliability threats. 

\textbf{Construct validity} is related to the connection between the research questions and the objects of study. 
In this respect, we used SonarQube to detect TD issues in our study and, thus, our interpretation of TD is limited to the capabilities of the tool. In particular, SonarQube defines several thresholds for some rules. For example, it will flag a new violation of the rule ``Lines should have sufficient coverage by tests'' when a test covers less than 65\% of the lines of code; and a violation of the rule ``Source files should have a sufficient density of comment lines'' is flagged for a file as soon as the density of comment lines on this file is less than 25\%.
Although SonarQube has been widely used in both industry and academia, tools using different strategies could lead to variations in the TD issues, and in turn, self-fixed issues. 
In principle, any used tool would be subject to similar threats.

There might also be potential threats due to false positives flagged by SonarQube. To assess SonarQube's limitations, we manually analyzed 0.5\% of the number of total fixed issues (i.e., $\approx 350$) to investigate whether those issues represent technical debt and whether their evolution is accurately captured. For that, we randomly selected issues by using \textit{stratified random sampling}~\cite{Stratified_sample}, which is used to estimate population parameters efficiently when subpopulations have substantial variability~\cite{sampling_techniques}. For each rule, we randomly selected a number of issues based on its fixing prevalence (i.e., the results shown in Table~\ref{tab:RQ2}) and checked their source code to verify whether they may represent TD according to the definitions of Alves et al.~\cite{ClassifyTD2014} and Li et al.~\cite{Li2015SMS}. Then, we checked if the issue was actually fixed. The results of our analysis show that all selected issues are potential TD issues (i.e., they match the used definitions). Moreover, only seven issues were not actually fixed.

There is also a threat associated with the method for filtering out issues that disappeared due to the file deletion, since it is doubtful whether developers actually aimed at fixing the technical debt. This may lead to the erroneous deletion of some issues that the developers did fix. Moreover, we cannot understand the developer's intent by analyzing source code. Filtering out such issues is a safe way to conducting our study. In the case of file renaming, SonarQube can still track the issues in the renamed file (if the issues still exist in the system).

Another threat pertains to how we analyze the evolution of projects. 
Due to the amount of computational effort, we limited the data collection to 2,000 commits per project, which affected five of them. 
The incomplete analysis of some of the projects may have led to missing aspects of TD evolution and involved developers. 
This threat is partially mitigated as, the majority of issues are likely to be fixed within one year~\cite{DigkasSANER2018}, and the data collected per project spans at least one year and almost four years in average. 

Moreover, although we considered four different factors (i.e., seniority, file count, commit count and commit size) for each developer, there are still other factors related to the likelihood of a developer self-fixing TD, such as the commit content and code comments. It is however, rather difficult to consider all of these factors in one study; thus we could not fully mitigate this threat.

Furthermore, mining git repositories presents potential risks. We restricted our study to the main branch only, as the investigation focuses on technical debt that has been repaid in the final product. However, different team strategies may affect how the main branch is used. The main threat in this approach is that we may be missing debt items indirectly relevant to the final product, e.g., feature branches that had debt introduced and repaid locally before merging into the main branch.
Related threats also include, for example, that fixing commits might be suppressed in parent commits (i.e., squashing), or the commit might be a pull request from a third party that is attributed wrongly. Although off-branches are out of the scope of the study, merging commits from other branches has the potential to obscure the actual author of the repayment depending on team strategies (e.g., defining who squashes commits and when). This threat can be challenging to mitigate without contacting the development team. To partially address it, we inspected a random sample of 20 (non-merge) commits from each project (approx. 1\%) to check whether the commit author appears to have also authored the code (e.g., through commit comments or forked repository). This inspection suggest that the authors of the verified commits are the actual code authors.

Other related threats could be the influence of pair or mob programming practices and code review on how technical debt is managed. Such quality improvement practices sometimes are not introduced in main branches but in short-lived branches. Specifically, these practices can reduce the amount of introduced debt or increase the amount of repaid debt. We looked for evidence of such practices in the documentation and website of the projects and mainly identified code review as a common practice, although we cannot know how it affected TD management (e.g., how much extra effort was dedicated to it).

\textbf{External validity} concerns threats to the generalizability of our findings. 
The main threats stem from our study design, which considers the evolution of 20 Python and 16 Java projects from the Apache Software Foundation. 
Despite the credibility of the Apache Foundation and the diversity of its Python and Java projects, our results may not fully represent the entire population of non-trivial Python and Java projects. In principle, our results can be generalized to large and complex Python and Java projects in ecosystems similar to Apache.
Furthermore, the set of rules considered in this study is not exhaustive. In fact, there is no complete set of issues related to TD that may affect Python or Java source code.

\textbf{Reliability} concerns the degree to which conducting the study depends on the involved researchers. 
To address these threats, three researchers were involved in the data collection and analysis. 
Moreover, samples of the analysis output from different steps were manually inspected for irregularities and for consistency with the proposed study design. The results showed no irregularity and all the output from different steps were consistent with the proposed study design. Furthermore, the first and second authors classified the rules independently into the five TD categories, using the description of the rules and the definition of the categories. To assess the disagreements numerically, we estimated the inter-rater agreement using Krippendorff's alpha~\cite{krippendorff} ($\alpha = 0.74$).\footnote{Krippendorff’s inspection of the tradeoffs between statistical techniques establishes that it is customary to require $\alpha\geq0.80$. However, where tentative conclusions are still acceptable, $\alpha\geq0.67$ is the lowest conceivable limit~\cite{krippendorff_alpha}.}
Finally, most steps were automated by scripts which are publicly available together with the collected dataset.

\textbf{Confounding factors} are variables that may affect the dependent variables without the knowledge of the researchers. In our study, the main limitations that we expect pertain to the differences between Python and Java. The most prominent factors are related to the characteristics of the projects. Although language features can affect the complexity of the project, and thus affect the final dataset and findings, the complexity of the source code can also be influenced by the type of project and developers' experience, which may also not be uniform among Java and Python projects.

\section{Associated Dataset and Replication}
\label{sec:Associated Dataset and Replication}

We created an online repository\footnote{\url{https://github.com/jieshanshan/ist-si-sftd}} with instructions and scripts to replicate the data collection and support the analyses performed in our study. 
We note that the data collection takes a long time. 
The analysis of the approx. 44K commits from the 36 Python and Java projects by SonarQube took approx. three months and a half of work using two personal computers, one with an Intel Core i7-5500U and 8GB of RAM and another with an Intel i7-8550U and 32GB of RAM.

In short, the replication package helps with setting up the necessary environment (i.e., tools and configuration) through a Vagrant script to bootstrap a virtual machine and automate most of the setup. 
The environment includes SonarQube 7.0, with a PostgreSQL database, and Jupyter\footnote{\url{https://jupyter.org/}}, which is used to support the data collection.

\section{Conclusions and Future Work}
\label{sec:Conclusion and Future Work}

This paper reports on an empirical study that investigated the phenomenon of self-fixed technical debt. 
We analyzed the evolution of TD in consecutive 17K commits of 20 Python projects and 27K commits of 16 Java projects from the Apache Software Foundation. 
Moreover, we detected 54K Python TD issues and 71K Java TD issues based on 49 Python rules and 52 Java rules of \textit{blocker}, \textit{critical} or \textit{major} severity levels, as defined by SonarQube. 

We found that about half of fixed TD issues were repaid by the same developers who introduced them. Moreover, the projects that are larger and have more commits and technical debt issues tend to have a relatively lower likelihood of TD issues being self-fixed. We also observed that there is no significant difference in the survival time between self-fixed and non-self-fixed issues.

Looking at the types of TD, our findings suggest that Defect Debt and Code Debt are more likely to be self-fixed than other debt types. The results indicate that developers mostly pay back technical debt that was introduced by them when it is related to lower code level issues (e.g., compared to Design Debt and Documentation Debt).

Finally, regarding developer-related factors, the results indicate that those who are more dedicated to or knowledgeable of the project (e.g., by contributing more commits) may drive the overall self-fixing up. Moreover, the longer the developers are involved in the project, the less often they self-fix TD. 

In the future, we plan to extend this study by investigating more factors related to the survival time and the likelihood of TD issues being self-fixed, such as commit frequency at project and developer levels, and the commit content. We also plan to develop a tool based on the self-fixed rate and survival time to help developers to prioritize TD remediation, and provide insights to software development teams to better assign software maintenance tasks among different developers.

\section*{Acknowledgements}

This work was supported by ITEA3 and RVO under grant agreement No. 17038 VISDOM (\url{https://visdom-project.github.io/website/}).

\bibliographystyle{elsarticle-num}
\bibliography{mybibfile}

\begin{appendices}

\newpage
\section{Self-fixing rates for individual rules}
\label{append:rules}
\setlength{\tabcolsep}{1pt}

\begin{center}
\footnotesize
 \setlength{\tabcolsep}{2pt} 
    \begin{longtable}{clp{10cm}rrrr}
        \caption{Self-fixing rates for individual rules}
        \label{tab:RQ2}\\
    \hline
    \textbf{Language} & \textbf{ID}$^{a}$ & \textbf{Definition} &
    \textbf{SR}$^{b}$ & \textbf{FR}$^{c}$ & \textbf{Issues} & \textbf{P}$^{d}$\\
    \hline
    \endfirsthead
    \hline
    \textbf{Language} & \textbf{ID}$^{a}$ & \textbf{Definition} &
    \textbf{SR}$^{b}$ & \textbf{FR}$^{c}$ & \textbf{Issues} & \textbf{P}$^{d}$\\
    \hline
    \endhead
    \hline
    \endfoot
    Python & \CodeDebt438 & Assigning to function call which doesn't return & 1.00  & 1.00  & 1     & 1 \\
    Python & \CodeDebt25 & Redundant pairs of parentheses should be removed & 1.00  & 0.13  & 8     & 2 \\
    Python & \DefectDebt368 & Not enough arguments for format string & 1.00  & 1.00  & 1     & 1 \\
    Python & \DefectDebt412 & Multiple values passed for parameter in function call & 1.00  & 1.00  & 1     & 1 \\
    Python & \CodeDebt43 & Related "if/else if" statements should not have the same condition & 1.00  & 1.00  & 1     & 1 \\
    Java  & \DefectDebt775 & Package declaration should match source file directory & 1.00  & 1.00  & 1557  & 4 \\
    Python & \DefectDebt393 & Using variable before assignment & 0.80  & 0.56  & 9     & 6 \\
    Python & \DefectDebt360 & Undefined name & 0.75  & 0.53  & 1675  & 16 \\
    Java  & \CodeDebt422 & Methods and field names should not be the same or differ only by capitalization & 0.73  & 0.30  & 275   & 14 \\
    Python & \DefectDebt432 & Mixed tabs/spaces indentation & 0.72  & 0.98  & 54    & 3 \\
    Python & \CodeDebt13 & Jump statements should not be followed by other statements & 0.71  & 0.83  & 29    & 6 \\
    Java  & \DefectDebt423 & Anonymous inner classes containing only one method should become lambdas & 0.67  & 0.40  & 718   & 12 \\
    Python & \DefectDebt49 & Variables should not be self-assigned & 0.67  & 0.27  & 11    & 5 \\
    Java  & \CodeDebt717 & Factory method injection should be used in "@Configuration" classes & 0.67  & 0.50  & 6     & 1 \\
    Java  & \DefectDebt677 & Encryption algorithms should be used with secure mode and padding scheme & 0.67  & 0.64  & 28    & 5 \\
    Python & \DefectDebt51 & Python parser failure & 0.65  & 1.00  & 31    & 9 \\
    Python & \DefectDebt328 & Redefined function/class/method & 0.60  & 0.78  & 127   & 11 \\
    Python & \CodeDebt7 & Methods and field names should not differ only by capitalization & 0.58  & 0.41  & 29    & 3 \\
    Java  & \DefectDebt288 & Asserts should not be used to check the parameters of a public method & 0.58  & 0.18  & 846   & 7 \\
    Java  & \CodeDebt285 & Try-with-resources should be used & 0.56  & 0.56  & 224   & 14 \\
    Python & \DefectDebt363 & Method should have "self" as first argument & 0.52  & 0.53  & 116   & 5 \\
    Python & \DesignDebt24 & Control flow statements "if". "for". "while". "try" and "with" should not be nested too deeply & 0.51  & 0.41  & 499   & 14 \\
    Python & \DefectDebt281 & Syntax error & 0.50  & 0.64  & 817   & 19 \\
    Python & \CodeDebt45 & Nested blocks of code should not be left empty & 0.50  & 0.55  & 51    & 11 \\
    Python & \DesignDebt41 & Collapsible "if" statements should be merged & 0.49  & 0.54  & 226   & 15 \\
    Java  & \CodeDebt632 & Constructors should not be used to instantiate "String", "BigInteger", "BigDecimal" and primitive-wrapper classes & 0.48  & 0.78  & 321   & 13 \\
    Python & \CodeDebt16 & Sections of code should not be "commented out" & 0.47  & 0.73  & 566   & 18 \\
    Python & \CodeDebt42 & Function names should comply with a naming convention & 0.47  & 0.46  & 321   & 15 \\
    Python & \DefectDebt21 & The "print" statement should not be used & 0.47  & 0.68  & 1478  & 18 \\
    Python & \CodeDebt38 & Statements should be on separate lines & 0.47  & 0.50  & 1282  & 7 \\
    Python & \DefectDebt418 & Too many positional arguments for function call & 0.46  & 0.24  & 110   & 6 \\
    Python & \CodeDebt36 & A field should not duplicate the name of its containing class & 0.46  & 0.46  & 28    & 6 \\
    Java  & \TestDebt425 & Tests should not be ignored & 0.46  & 0.26  & 360   & 11 \\
    Java  & \DesignDebt573 & Two branches in a conditional structure should not have exactly the same implementation & 0.44  & 0.50  & 214   & 15 \\
    Java  & \DefectDebt476 & Raw types should not be used & 0.44  & 0.51  & 8871  & 16 \\
    Python & \DefectDebt403 & Too few arguments & 0.43  & 0.93  & 57    & 9 \\
    Java  & \CodeDebt486 & Unused "private" classes should be removed & 0.42  & 0.67  & 36    & 13 \\
    Python & \DefectDebt276 & Access of nonexistent member & 0.42  & 0.58  & 1871  & 19 \\
    Python & \DesignDebt29 & Cognitive Complexity of functions should not be too high & 0.41  & 0.50  & 1353  & 19 \\
    Java  & \DefectDebt289 & Resources should be closed & 0.41  & 0.49  & 385   & 15 \\
    Java  & \DesignDebt779 & Cognitive Complexity of methods should not be too high & 0.40  & 0.45  & 4703  & 16 \\
    Python & \DesignDebt390 & Method hidden by attribute of super class & 0.40  & 0.80  & 25    & 3 \\
    Java  & \CodeDebt431 & Sections of code should not be commented out & 0.40  & 0.64  & 5444  & 16 \\
    Python & \DefectDebt394 & Undefined variable & 0.40  & 0.62  & 4361  & 18 \\
    Java  & \CodeDebt369 & Constant names should comply with a naming convention & 0.39  & 0.43  & 1441  & 14 \\
    Java  & \CodeDebt784 & Synchronized classes Vector, Hashtable, Stack and StringBuffer should not be used & 0.38  & 0.54  & 551   & 16 \\
    Java  & \CodeDebt437 & switch statements should have "default" clauses & 0.37  & 0.26  & 596   & 14 \\
    Python & \DesignDebt26 & Two branches in a conditional structure should not have exactly the same implementation & 0.37  & 0.65  & 195   & 15 \\
    Java  & \CodeDebt389 & Standard outputs should not be used directly to log anything & 0.37  & 0.28  & 2735  & 16 \\
    Java  & \DefectDebt453 & ThreadLocal variables should be cleaned up when no longer used & 0.37  & 0.41  & 140   & 14 \\
    Java  & \CodeDebt799 & Unused "private" methods should be removed & 0.37  & 0.60  & 472   & 15 \\
    Java  & \DefectDebt750 & Local variables should not shadow class fields & 0.37  & 0.48  & 1404  & 16 \\
    Java  & \CodeDebt386 & Nested blocks of code should not be left empty & 0.36  & 0.61  & 522   & 15 \\
    Java  & \CodeDebt681 & Unused "private" fields should be removed & 0.36  & 0.68  & 793   & 16 \\
    Java  & \DesignDebt337 & String literals should not be duplicated & 0.35  & 0.48  & 3343  & 16 \\
    Java  & \DesignDebt837 & Source files should not have any duplicated blocks & 0.35  & 0.46  & 3121  & 16 \\
    Python & \DefectDebt424 & NotImplemented raised - should raise NotImplementedError & 0.35  & 0.96  & 74    & 4 \\
    Python & \DocumentationDebt57 & Source files should have a sufficient density of comment lines & 0.35  & 0.52  & 2004  & 20 \\
    Python & \DefectDebt274 & Calling of not callable & 0.35  & 0.72  & 115   & 7 \\
    Python & \CodeDebt46 & Functions. methods and lambdas should not have too many parameters & 0.35  & 0.75  & 729   & 12 \\
    Java  & \CodeDebt693 & Unused method parameters should be removed & 0.35  & 0.55  & 2988  & 16 \\
    Python & \DesignDebt40 & Functions should not contain too many return statements & 0.34  & 0.39  & 344   & 15 \\
    Python & \DefectDebt355 & Method has no argument & 0.33  & 0.50  & 36    & 7 \\
    Python & \DefectDebt411 & Bad first argument given to super & 0.33  & 0.63  & 24    & 7 \\
    Java  & \CodeDebt596 & entrySet() should be iterated when both the key and value are needed & 0.33  & 0.36  & 161   & 14 \\
    Java  & \TestDebt434 & Thread.sleep should not be used in tests & 0.33  & 0.13  & 1357  & 16 \\
    Java  & \DocumentationDebt730 & Deprecated elements should have both the annotation and the Javadoc tag & 0.32  & 0.58  & 1214  & 16 \\
    Java  & \DefectDebt323 & Class names should not shadow interfaces or superclasses & 0.32  & 0.26  & 157   & 11 \\
    Python & \DesignDebt58 & Source files should not have any duplicated blocks & 0.31  & 0.63  & 604   & 17 \\
    Java  & \CodeDebt753 & Assertion arguments should be passed in the correct order & 0.31  & 0.40  & 1597  & 15 \\
    Java  & \CodeDebt624 & Empty arrays and collections should be returned instead of null & 0.30  & 0.34  & 506   & 16 \\
    Java  & \DefectDebt405 & Generic exceptions should never be thrown & 0.30  & 0.44  & 4695  & 16 \\
    Java  & \CodeDebt333 & String function use should be optimized for single characters & 0.30  & 0.41  & 235   & 16 \\
    Java  & \CodeDebt754 & Utility classes should not have public constructors & 0.29  & 0.23  & 1177  & 16 \\
    Java  & \DesignDebt671 & Collapsible "if" statements should be merged & 0.29  & 0.51  & 1343  & 16 \\
    Python & \DocumentationDebt14 & Docstrings should be defined & 0.28  & 0.54  & 13109 & 20 \\
    Python & \DesignDebt20 & Functions should not be too complex & 0.28  & 0.53  & 328   & 13 \\
    Java  & \DefectDebt666 & Methods should not be empty & 0.27  & 0.29  & 3367  & 16 \\
    Java  & \DefectDebt542 & Generic wildcard types should not be used in return parameters & 0.27  & 0.47  & 1306  & 15 \\
    Python & \CodeDebt50 & Lines should not be too long & 0.26  & 0.85  & 16168 & 19 \\
    Java  & \CodeDebt752 & Unused type parameters should be removed & 0.25  & 0.78  & 892   & 13 \\
    Python & \CodeDebt23 & \textbackslash{} should only be used as an escape character outside of raw strings & 0.25  & 0.86  & 1423  & 8 \\
    Java  & \TestDebt740 & Tests should include assertions & 0.24  & 0.27  & 2578  & 16 \\
    Python & \DefectDebt414 & Passing unexpected keyword argument in function call & 0.23  & 0.85  & 46    & 5 \\
    Python & \CodeDebt15 & Identical expressions should not be used on both sides of a binary operator & 0.23  & 0.84  & 31    & 3 \\
    Python & \DefectDebt369 & Too many arguments for format string & 0.23  & 0.94  & 33    & 4 \\
    Java  & \DocumentationDebt601 & @Override should be used on overriding and implementing methods & 0.22  & 0.66  & 1419  & 16 \\
    Python & \TestDebt56 & Lines should have sufficient coverage by tests & 0.21  & 0.37  & 2683  & 20 \\
    Python & \DesignDebt47 & Files should not have too many lines of code & 0.20  & 0.57  & 157   & 9 \\
    Java  & \DefectDebt828 & Boolean expressions should not be gratuitous & 0.19  & 0.59  & 472   & 15 \\
    Python & \DefectDebt439 & Raising only allowed for classes. instances or strings & 0.18  & 0.10  & 110   & 3 \\
    Java  & \CodeDebt341 & Static fields should not be updated in constructors & 0.18  & 0.51  & 43    & 9 \\
    Java  & \DefectDebt651 & Printf-style format strings should be used correctly & 0.18  & 0.45  & 1391  & 15 \\
    Python & \DefectDebt389 & Access to member before its definition & 0.18  & 0.63  & 62    & 5 \\
    Java  & \CodeDebt447 & Unused assignments should be removed & 0.17  & 0.70  & 2448  & 15 \\
    Java  & \DefectDebt819 & Overrides should match their parent class methods in synchronization & 0.14  & 0.29  & 73    & 11 \\
    Java  & \DefectDebt607 & Exceptions should be either logged or rethrown but not both & 0.13  & 0.10  & 515   & 10 \\
    Java  & \CodeDebt595 & Preconditions and logging arguments should not require evaluation & 0.13  & 0.45  & 919   & 15 \\
    Java  & \DefectDebt464 & Fields in a "Serializable" class should either be transient or serializable & 0.10  & 0.55  & 1083  & 16 \\
    Java  & \CodeDebt758 & static members should be accessed statically & 0.10  & 0.55  & 164   & 12 \\
    Java  & \DefectDebt391 & readResolve methods should be inheritable & 0.05  & 0.61  & 33    & 3 \\
    \hline
    \multicolumn{7}{l}{$^{a}$ ID number of the rule and Debt Type} \\
    \multicolumn{7}{l}{(\CodeDebt =Code Debt, \DesignDebt = Design Debt, \DocumentationDebt =Documentation Debt, \DefectDebt =Defect Debt, \TestDebt =Test Debt)}\\
    \multicolumn{7}{l}{$^{b}$ Self-fixing Rate: the ratio is the number of self-fixed issues divided by the number of all fixed issues}\\
    \multicolumn{7}{l}{$^{c}$ Fixing Rate: the ratio is the number of fixed issues divided by the number of total issues}\\
    \multicolumn{7}{l}{$^{d}$ The number of projects that contain the issues}\\
\end{longtable}
\end{center}

\end{appendices}

\end{document}